\documentclass[useAMS,usenatbib]{mn2e}
\usepackage{graphicx}

\title[Optical counterparts of two ULXs]
{Optical counterparts of two ULXs in NGC\,5474 and NGC\,3627 (M\,66)}
\author[S. Avdan et al.]
  {S.~Avdan,$^{1,2}$\thanks{kayaci.s@gmail.com.}
   A.~Vinokurov,$^3$ S.~Fabrika,$^{3,4}$ K.~Atapin,$^{3, 5}$ H.~Avdan,$^{1,2}$ A.~Akyuz,$^{1,2}$ 
   \newauthor
    O.~Sholukhova,$^3$ N. Aksaker,$^{2,6}$ A.~Valeev$^{3,4}$\\
  $^1$Department of Physics, University of Cukurova, Adana 01330, Turkey\\
  $^2$Space Sciences and Solar Energy Research and Application Center (UZAYMER), University of Cukurova, Adana 01330, Turkey \\
  $^3$Special Astrophysical Observatory of the Russian AS, Nizhnij Arkhyz 369167, Russia\\
  $^4$Kazan Federal University, Kazan 420008, Russia\\
  $^5$Sternberg Astronomical Institute, Moscow State University 119991, Russia\\
  $^6$Vocational School of Technical Sciences, University of Cukurova, Adana 01410, Turkey\\
  }
\date{Released 2002 Xxxxx XX}

\pagerange{\pageref{firstpage}--\pageref{lastpage}} \pubyear{2002}

\def\LaTeX{L\kern-.36em\raise.3ex\hbox{a}\kern-.15em
    T\kern-.1667em\lower.7ex\hbox{E}\kern-.125emX}

\begin{document}

\label{firstpage}

\maketitle

\begin{abstract}
We identified two optical counterparts of brightest ultraluminous X-ray Sources (ULXs) in galaxies NGC\,5474 and NGC\,3627 (M66). The counterparts in {\it Hubble Space Telescope} images are very faint, their $V$ magnitudes are 24.7 ($M_V \approx -4.5$) and 25.9 ($M_V \approx -4.2$), respectively. NGC\,5474 X-1 changes the X-ray flux more than two orders of magnitude, in its bright state it has $L_{\mathrm{X}} \approx 1.6 \times10^{40}$\,erg s$^{-1}$, the spectrum is best fitted by an absorbed power-law model with a photon index $\Gamma \approx 0.94$. M66 X-1 varies in X-rays with a factor of $\sim 2.5$, its maximal luminosity being $2.0 \times 10^{40}$ erg s$^{-1}$ with $\Gamma \approx 1.7$. Optical spectroscopy of the NGC\,5474 X-1 has shown a blue spectrum, which however was contaminated by a nearby star of 23 mag, but the counterpart has a redder spectrum. Among other objects captured by the slit are a background emission-line galaxy (z $= 0.359$) and a new young cluster of NGC\,5474. We find that these two ULXs have largest X-ray-to-optical ratios of $L_{\mathrm{X}}/L_{\mathrm{opt}} \sim 7000$ for NGC\,5474 X-1 (in its bright state) and 8000 for M66 X-1 both with the faintest optical counterparts ever measured. Probably their optical emission originates from the donor star. If they have super-Eddington accretion discs with stellar-mass black holes, they may also have the lowest mass accretion rates among ULXs such as in M81 X-6 and NGC\,1313 X-1.
\end{abstract}


\section{Introduction}

Ultraluminous X-ray sources (ULXs) are sources with X-ray luminosities exceeding the Eddington limit for a typical stellar-mass black hole \citep{Feng2011}, $\sim 2 \times 10^{39}$ erg s$^{-1}$. The most popular models of ULXs involve either intermediate-mass black holes (IMBHs) of $10^2$$-$$10^4 M_{\odot}$ with standard accretion discs \citep{Colbert1999} or stellar-mass black holes ($\sim 10 M_{\odot}$) accreting at super-Eddington rates \citep{Poutanen2007}. However, both scenarios require a massive donor in a close binary.

Most ULXs are associated with the star-forming regions (Swartz et al. 2009). They are not distributed throughout galaxies as it would be expected for IMBHs originating from low-metallicity Population III stars \citep{Madau2001}. The IMBHs may be produced in  runaway mergers in the cores of young clusters \citep{Portegies2004, Freitag2006}. In such cases, they usually  stay within their clusters. It has been found \citep{Poutanen2013} that all brightest X-ray sources in the Antennae galaxies are located nearby the very young stellar clusters. They concluded that these sources were ejected in the process of formation of stellar clusters in the dynamical few-body encounters and that the majority of ULXs are massive X-ray binaries with the progenitor masses larger than $50~M_{\odot}$.

The X-ray spectra of the ULXs often show a high-energy curvature \citep{Stobbart2006, Gladstone2009, Caballero2010} with a downturn between $\sim 4$ and $\sim 7$\,keV. The curvature hints that the ULX accretion discs are no standard. Inner parts of the discs may be covered with a hot outflow or optically thick corona \citep{Gladstone2009}, Comptonizing the inner disc photons.

Multiwavelength study of ULXs is a key issue to investigate their nature and environment. Optical investigations show that some ULXs are associated with optical emission nebulae \citep{Abolmasov2007a, Feng2011}. Optical counterparts of several ULXs have been identified with {\it Hubble Space Telescope} ({\it HST}) in nearby galaxies, for example NGC\,4559 \citep{Soria2005}, NGC\,5204 \citep{Liu2004} and Holmberg~II \citep{Kaaret2004}. In addition, optical spectra of some ULX counterparts and their environments also have been obtained with various ground-based telescopes and {\it HST}: ULXs in M101 \citep{Kuntz2005}, NGC 7331 \citep{Abolmasov2007b}, Holmberg IX \citep{Grise2011} and  NGC 7793 \citep{Motch2014}. On the other hand, some ULXs have no visible counterparts even in {\it HST} images, for example NGC\,5128 X-1 or some sources in NGC\,3034 \citep{Gladstone2013}.        

In the recent spectroscopy of several optical ULX counterparts with {\it Subaru} telescope, \cite{Fabrika2015} have found that all these spectra, and also practically all previously published optical spectra of ULXs, are similar to one another and also to the optical spectrum of SS\,433 \citep{Fabrika1997, Fabrika2004}. SS\,433 is the only known supercritical accretor in our Galaxy, a close binary with a stellar-mass black hole. The spectra contain broad emission lines of He\,II $\lambda 4686$ and that of hydrogen (H$\alpha$ and H$\beta$) with full width at half-maximum (FWHM) $\sim 1000$~km\,s$^{-1}$. In X-rays, SS\,433 is very different \citep{Medvedev2010} from ULXs, both in the spectrum and X-ray luminosity. Nevertheless, it was suggested that being observed nearly face-on, the supercritical accretion disc in SS\,433 is expected to be extremely bright X-ray source \citep{Fabrika2004}.

Here, we present optical identification of two new ULXs in galaxies NGC\,5474 and NGC\,3627 (M66). The first source is located in a star-forming region of the galaxy classified as a late-type dwarf spiral at a distance of 6.8 Mpc \citep{Drozdovsky2000}. The source NGC\,5474 X-1 is identified as a ULX by \cite{Swartz2011}, with an X-ray luminosity of $L_{\mathrm{X}} \simeq 1.4 \times$ $10^{40}$ erg s$^{-1}$. The second ULX is located on the edge of the disc of M66, which is an SAb-type barred galaxy at a distance of 10.6 Mpc \citep{Lee2013}. The ULX in this galaxy has luminosity of $L_{\mathrm{X}} \simeq$ 1.8 $\times$ $10^{40}$ erg s$^{-1}$ at the adopted distance.

Here, we also present X-ray and optical analysis of {\it Chandra} and {\it HST} observations of these two sources including the optical spectroscopy of the region surrounding NGC\,5474 X-1 taken with the 6 m BTA (Big Telescope Alt-azimuth) telescope and discuss the X-ray and optical luminosities of these two sources in comparison with other well-known ULXs. 

\section{Observations and Analysis}

NGC 5474 was observed twice by {\it Chandra} ACIS-S on  2006 September 10 for $\sim$ 2 ks (ObsID 7086) and 2007 December 03 for $\sim$ 30 ks (ObsID 9546). {\it Chandra} observations of M66 were performed on 1999 November 3 for $\sim$ 2 ks (ObsID 394) and 2008 March 31 for $\sim$ 50 ks (ObsID 9548). For M66, there are also the {\it XMM--Newton} data (ObsID 0093641101) obtained on 2001 May 26. To test the long-time variability of the sources we used all available observations obtained with {\it Swift}. There are 14 {\it Swift} data sets for NGC\,5474. The first one was obtained on 2011 April 15 with an exposure time of 5 ks. Other 13 observations were obtained from 2012 June 20 to November 21 with a total exposure of 13 ks, there were 0-2 counts in each of individual observation. M66 was observed by {\it Swift} four times, from 2008 July 20 to 2013 November 1. 

In Fig.\,1 we present light curves of these two sources. NGC\,5474 X-1 shows a strong variability. Its luminosity changes from $1.6 \times 10^{40}$ to $\la 10^{38}$\,erg s$^{-1}$. During the last 13 {\it Swift} observations, its average luminosity was about $\sim 2.5 \times 10^{38}$\,erg s$^{-1}$. For all {\it Swift} data, we used 30 arcsec circular aperture (Fig.\,1). In the later 13 observations of NGC\,5474, the net count rate was $6.5\times 10^{-4}$ s$^{-1}$ (orange circle in the figure), whereas the background was $5.3\times 10^{-4}$ s$^{-1}$. When we use 12 arcsec aperture, we find the net count rate $7.3\times 10^{-4}$ s$^{-1}$ and the background $0.9\times 10^{-4}$ s$^{-1}$. Therefore, we have detected the source in the last {\it Swift} observations. With the total variability amplitude more than two orders of magnitude, we may consider NGC\,5474 X-1 as a transient. On the other hand, M66 X-1 shows a variability with a factor of $\sim 2.5$ which is rather normal for ULXs (Feng \& Soria 2011). For M66 X-1, we use 12 arcsec aperture because of nearby sources around.

{\it Chandra} data reductions were carried out using {\scshape ciao} version 4.6 with the CALDB (Calibration Data base) version 4.5.9. All data were filtered in the 0.3$-$10~keV energy range. The spectra were grouped to have at least 20 counts per bin for good statistical quality. The spectral analysis was performed using {\scshape xspec} version 12.8.

Using the longest exposure observation, we find that NGC\,5474 X-1 has a hard X-ray spectrum. This spectrum is best fitted by an absorbed power-law model {\scshape phabs*pl}, giving a reduced $\chi^{2} =1.12$ with 189 degrees of freedom (dof). The best-fitting model has a power-law photon index of $\Gamma$ = $0.94\pm{0.06}$ and a hydrogen column density of $N_{H}$ = ($0.04\pm{0.02}$) $\times$ $10^{22}$ cm$^{-2}$. The calculated unabsorbed flux in the (0.3$-$10) keV energy range is $2.91\times 10^{-12}$ erg cm$^{-2}$ s$^{-1}$. We have found that the luminosity of the source is $L_{\mathrm{X}} \approx 1.6 \times$ $10^{40}$ erg s$^{-1}$. 

The X-ray spectrum of the ULX in M66 is best fitted with the same model at reduced $\chi^{2} \sim$ 1.35 with 184 dof (the longest exposured observation). The resulting model parameters are $\Gamma$ = $1.66\pm{0.06}$ and $N_{H}$ = ($0.14\pm{0.02}$) $\times$ $10^{22}$ cm$^{-2}$. The source has a flux of $1.46\times 10^{-12}$ erg cm$^{-2}$ s$^{-1}$. This yield a luminosity of $L_{\mathrm{X}}$=2.0 $\times$ $10^{40}$ erg s$^{-1}$.

\begin{figure*}
\label{Fig1}
\begin{center}
\includegraphics[angle=-90,scale=0.40]{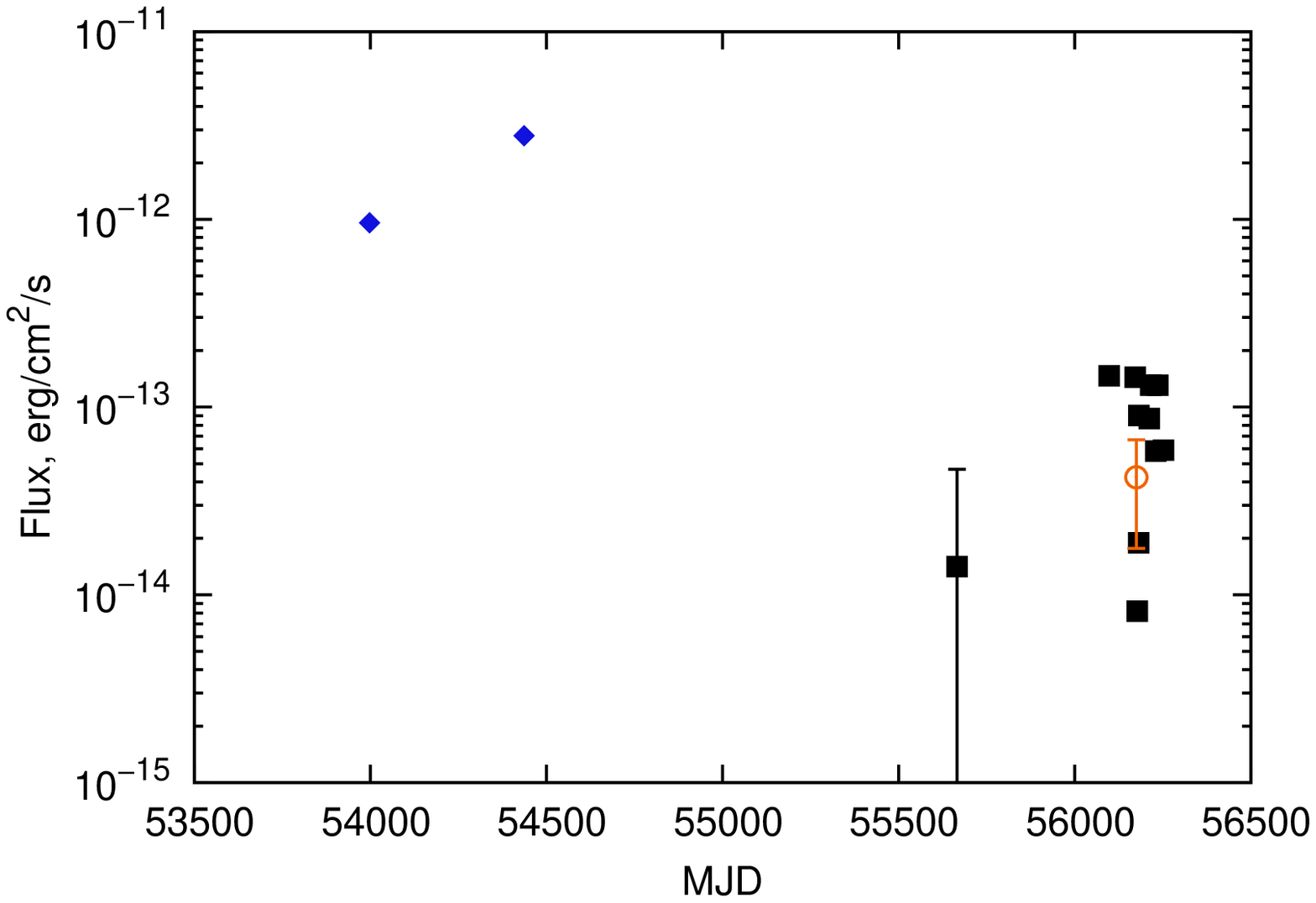}
\includegraphics[angle=-90,scale=0.40]{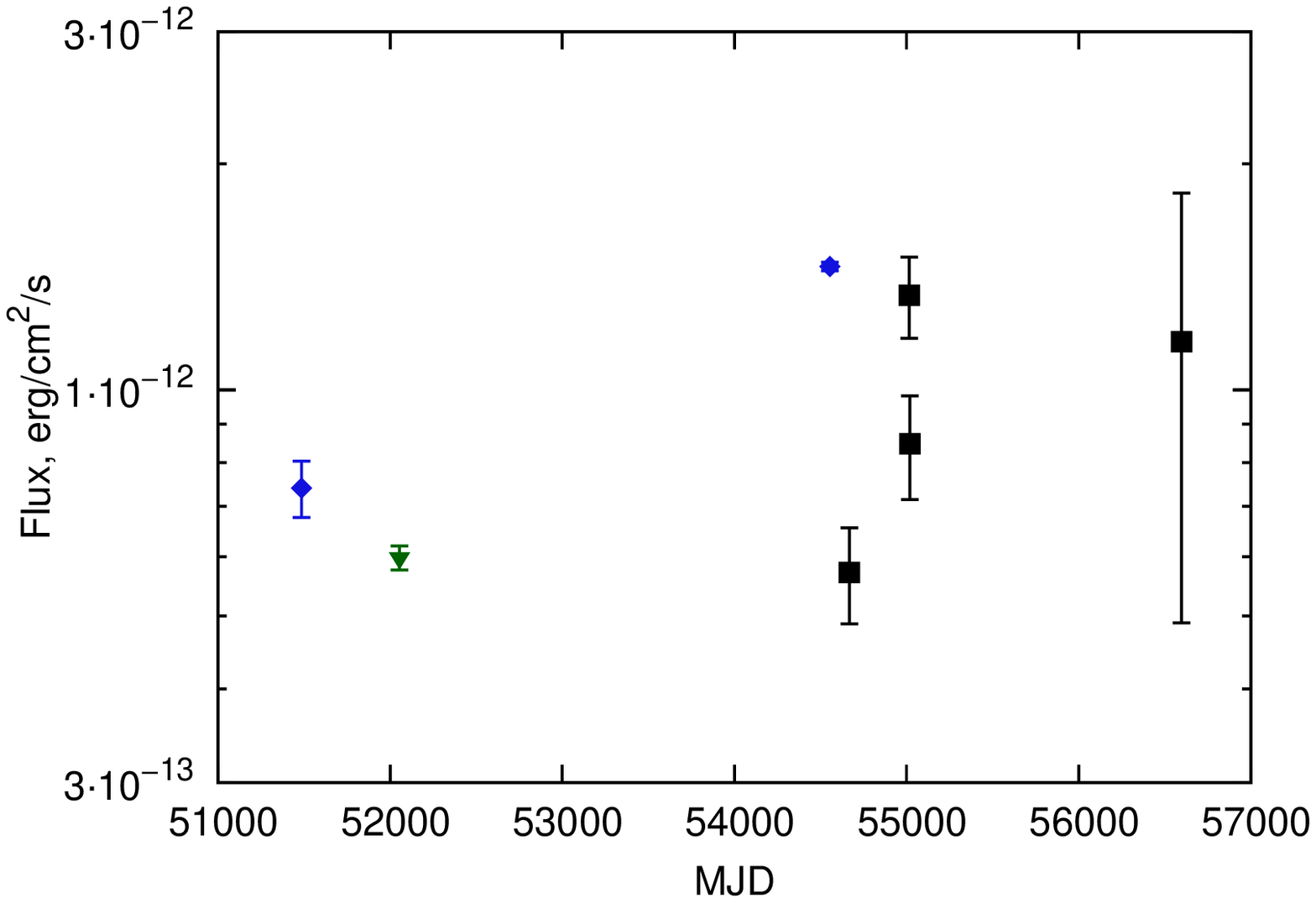}
\caption{The light curves of NGC\,5474 X-1 (left) and M66 X-1 (right) taken with {\it Chandra} (blue diamonds), {\it XMM Newton} (green triangle) and {\it Swift} (black squares) in the range 0.3$-$10 keV. The orange circle is an average of 13 individual {\it Swift} observations (some of them have zero counts). To estimate fluxes from {\it Swift} counts we used model parameters from the longest {\it Chandra} observations.}
\end{center}
\end{figure*}

To make astrometry and photometry of NGC\,5474, X-1 we used {\it HST} ACS/WFC/$F606W$ and $F814W$ images taken on 2012 February 26; for M66 X-1 we used ACS/WFC/$F435W$ image taken on 2004 December 31, ACS/WFC/$F555W$ and $F814W$ images taken on 2009 December 14. For X-ray images we used the longest {\it Chandra} observations for both sources. In each {\it Chandra} image, we identified six point-like sources with a flux larger than 30 counts which makes their identification unambiguous. In the result, we derived a position for NGC\,5474 X-1 RA = $14^{h}$ $04^{m}$ $59^{s}.746$, Dec. = $+$ $53^{\circ}$ $38\arcmin$ $08\arcsec.86$ with an accuracy better than 0.16 arcsec and a position for M66 X-1 RA = $11^{h}$ $20^{m}$ $20^{s}.910$, Dec. = $+$ $12^{\circ}$ $58\arcmin$ $46\arcsec.57$ with an accuracy better than 0.13 arcsec. In Fig.\,2, we present these positions of our ULXs.

\begin{figure*}
\label{Fig2}
\begin{center}
\includegraphics[scale=0.25]{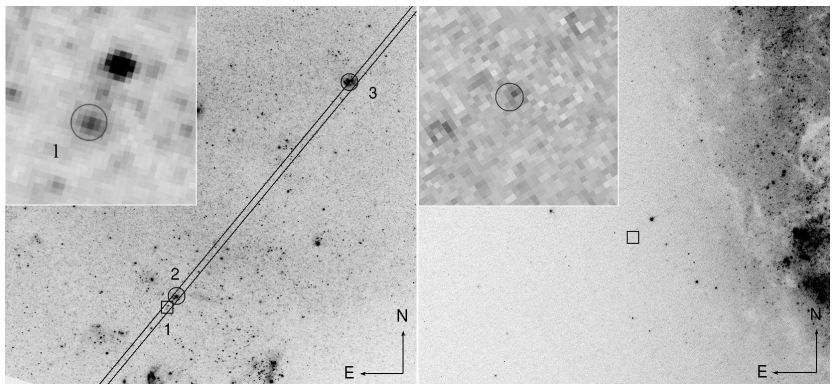}
\caption{The {\it Chandra} positions of the NGC\,5474 X-1 (left) and M66 X-1 (right) overplotted (square regions) on the astrometrical-corrected {\it HST} images, ACS/WFC/$F606W$ for NGC\,5474 and ACS/WFC/$F555W$ for M66. In the zoomed images, the circles of 0.16 arcsec radius (left) and 0.13 arcsec (right) indicate the errors. A length of the NE arrows is of 6.5 arcsec. In the NGC\,5474 image, we show the 1 arcsec slit (see Fig.\,3) capturing the ULX region (1), emission-line galaxy (2) and young star cluster (3).}
\end{center}
\end{figure*}

The photometry was performed on drizzled images, using the {\scshape apphot} package in {\scshape iraf}. All magnitudes are given in Vegamag system. The background was estimated from a concentric annulus around the objects. The optical counterpart of NGC\,5474 X-1 is extended (Fig.\,2), its size $\simeq$ 0.22 arcsec $\times$ 0.17 arcsec in the ACS/WFC/$F606W$ image, whereas the surrounding stars have FWHM $\simeq $\,0.11 arcsec. The counterpart may be complex. The measurements were performed with the aperture radius of 0.175 arcsec (3.5 pixels) in order to reduce contamination from nearby sources. The aperture corrections were not used due to a complex morphology of this object. The reddening correction was done in Synphot package using the Galactic extinction value $E(B - V) = 0.010$ \citep{Schlegel1998} for NGC\,5474. An extinction found in our spectra (see below), coincides with the Galactic value. The dereddened magnitudes of the ULX counterpart are $m_{F606W} = 24.58\,\pm\,0.04$ and $m_{F814W}=24.10\,\pm\,0.05$.

We have also carried out a photometry of the nearby objects. The brightest star shown in the zoomed image of Fig.\,2 (left), separated from the ULX counterpart by about 0.5 arcsec, has $m_{F606W}= 23.01\,\pm\,0.03$ and $m_{F814W}=23.09\,\pm\,0.02$. The magnitudes were corrected for aperture and reddening. This star is a major contributor to the spectrum of the object 1 (Figs 2 and 3). We also measured magnitudes of an emission-line galaxy (object 2 in the same figure), $m_{F606W}=21.828\,\pm\,0.009$ and $m_{F814W}=20.801\,\pm\,0.008$. The photometry of the star and the galaxy have been performed with aperture radii of 0.1 and 0.3 arcsec, respectively. 

\begin{figure}
\label{Fig3}
\begin{center}
\includegraphics[scale=0.100]{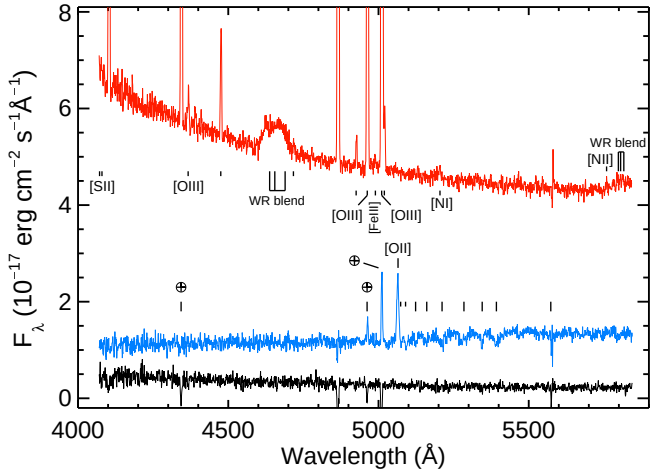}
\caption{Optical spectra of the NGC\,5474 X-1 region. From top to bottom: the young star cluster, the background galaxy and the ULX with nearby contaminating star numbered in Fig.\,2 as (3), (2) and (1) respectively. We have shifted the background galaxy spectrum by $+0.5\,\times\,10^{-17}$ and the cluster spectrum by $+3.5\,\times\,10^{-17}$ erg cm$^{-2}$ s$^{-1}$ \AA$^{-1}$. The spectrum (1) is blue with oversubtracted nebula emissions. The galaxy spectrum (2) contains [OII]~$\lambda$3727 emission line, absorption lines of Fe\,I $\lambda \lambda 3734, 3746$ (short vertical bars) and H$_{11}$, H$_{10}$, H$_{9}$, H$_{8}$, Ca\,II~K, H$\epsilon$, H$\delta$ (long vertical bars), all at a redshift of z$=$0.359. Two [OIII] lines and H$\gamma$ caused by the contamination of the host galaxy nebulae are marked with $\oplus$. The cluster spectrum (3) contains well-known WR blends and narrow hydrogen and forbidden lines. He\,I lines are shown by vertical bars.}
\end{center}
\end{figure}

The corrected position of M66 X-1 on the {\it HST} image can be seen within the X-ray error box in Fig.\,2. It looks as a single star with extended background in the $F435W$ and $F555W$ filters. In the $F814W$ filter however, in a southern part of the box, the background becomes brighter, and a redder object appears in the error box. Therefore, we choose two apertures, a smaller one (with a radius of 0.1 arcsec) for bright blue component and a larger one (0.2 arcsec) to measure the total flux without aperture corrections. A reddening correction for M66 has been done using the Galactic value $E(B - V) = 0.034$ (Schlegel et al. 1998). In the smaller aperture, we found the magnitudes $m_{F435W}=27.12\,\pm\,0.15$, m$_{F555W}=26.36\,\pm\,0.12$, and for $m_{F814W}=25.49\,\pm\,0.09$. In the larger aperture we found $m_{F435W}=26.60\,\pm\,0.18$, $m_{F555W}=25.94\,\pm\,0.15$ and $m_{F814W}=24.77\,\pm\,0.09$. The redder object contaminates the larger aperture.

Long-slit spectroscopy of the NGC\,5474 X-1 was obtained using the 6 m BTA telescope on 2014 January 02. The SCORPIO instrument was used with the grism VPHG1200G which covers a spectral range of 4000--5700~\AA\ and provides a resolution of 5~\AA\ \citep{Afanasiev2005}. The seeing was $\approx 1$ arcsec during the night. A total exposure of 5520 s was acquired. All the data reduction and calibrations were performed with {\scshape midas} procedures.

The optical spectra are shown in Fig.\,3. The ULX counterpart is strongly contaminated by the nearby star of 23 mag. We find that the spectrum is blue. We also determined extinction from the H$\gamma$/H$\beta$ line ratio using surrounding H II nebulae. We have found the extinction value for NGC 5474 as $E(B-V)\,=\,0.0\,\pm\,0.1$, therefore we exclude extinction contamination of the galaxy. Other objects captured by the slit are also shown in the figure. The cluster (3 in Fig.\,2) with z$=$0.00075 belongs to NGC\,5474. It shows emission lines typical for young clusters \citep{Schaerer1999}.

\subsection{Discussion and Conclusions}

We calculated a ratio of optical ($V$) to X-ray (0.3$-$10 keV) luminosities for our two ULXs; $L_{\mathrm{X}}/L_{\mathrm{opt}} = 110 \pm 40$ for NGC\,5474 X-1 and $L_{\mathrm{X}}/L_{\mathrm{opt}} = 8000 \pm 2000$ for M66 X-1. Optical luminosity of the NGC\,5474 counterpart has been derived by the extrapolation to the $V$ band. The X-ray luminosity has been taken from {\it Swift} data (orange circle in Fig.\,2) as the nearest date to the optical data (MJD $= 55983.9$). For M66 X-1, we used the $V$ band magnitude taken in 0.2 arcsec aperture and the nearest {\it Swift} observation (2009 July 4 and 7).

We have estimated the same ratio values for other well-studied ULXs: NGC\,6946 ULX-1, Holmberg\,IX X-1, Holmberg\,II X-1, NGC\,5408 X-1, IC342 X-1, NGC\,1313 X-1, NGC\,1313 X-2, NGC\,5204 X-1, NGC\,4559 X-7 and M81 X-6 \citep{Tao2011, Yang2011, Vinokurov2013, Sutton2013, Pintore2014}. The average value for these sources is $L_{\mathrm{X}}/L_{\mathrm{opt}} \sim 1600$, where a minimal one (260) is in NGC\,6946 ULX-1 and a maximal one (4200) in M81 X-6. All the X-ray luminosities were estimated in the 0.3$-$10 keV range. A well-known HLX-1 ESO243-49 \citep{Soria2012} has $L_{\mathrm{X}}/L_{\mathrm{opt}} \sim 1000$. Note that this ratio was derived for two different states of the HLX-1, when the luminosities were differed by a factor of 2. In the brighter state the ratio $L_{\mathrm{X}}/L_{\mathrm{opt}}$ is about 40\,\% higher than that in the fainter state. 

We find that the both our sources are outside of the ULX interval for $L_{\mathrm{X}}/L_{\mathrm{opt}}$: NGC\,5474 X-1 is below the range while M66 X-1 has a greater $L_{\mathrm{X}}/L_{\mathrm{opt}}$. If we compare the sources with AGNs, we find that absolute majority of AGNs have the ratio from 0.1 to 10 \citep{Aird2010}, in extreme cases of heavily obscured AGNs the ratio is approached $\sim 100$ \citep{Della Ceca2015}. We may conclude that if a source has $L_{\mathrm{X}}/L_{\mathrm{opt}} \ga 100$, it certainly must be a ULX. We do not have an optical luminosity of NGC\,5474 X-1 in its X-ray bright state. If its counterpart does not change notably the optical brightness (as it is in other  ULX counterparts), in the bright state the ratio will be $L_{\mathrm{X}}/L_{\mathrm{opt}} \sim 7000$. AGNs can not change their X-ray luminosities more than two orders of magnitude as we observed in NGC\,5474 X-1. Therefore, this source is obviously a ULX.

It was suggested from X-ray and optical data \citep{Sutton2013, Fabrika2015} that ULXs are super-Eddington sources with stellar-mass black holes. If in standard accretion discs, the total luminosity is scaled with the mass accretion rate as $L \propto \dot M_0$, in the supercritical discs, it depends on black hole mass and is nearly independent of mass accretion rate, $L \propto L_{\rm Edd} (1 + a \ln(\dot M_0 / \dot M_{\rm Edd}))$, where $L_{\rm Edd} \approx 1.5 \times 10^{39} m_{10}$\,erg s$^{-1}$ is the Eddington luminosity, and $\dot M_{\rm Edd}$ corresponding mass accretion rate, $m_{10}$ is a black hole mass expressed in $10 M_{\odot}$, and $a \sim 0.5-0.7$ is the parameter accounting for advection (Poutanen et al. 2007). This is because the excess gas is expelled as a disc wind and the accreted gas is advected with the photon trapping, contributing little to the photon luminosity. The mass accretion rate can provide a factor of several in the logarithmic term. Besides that, the funnel in the super-Eddington disc wind will collimate the X-ray radiation to an observer also with a factor of a few, when observer may look at the funnel \citep{Ohsuga2011}. Thus, the apparent X-ray luminosity of a ULX with a stellar mass black hole may well be up to $\sim 10^{41}$ erg s$^{-1}$.

In the super-Eddington disc the ultraviolet/optical luminosity does depend stronger on the mass accretion rate, because the excess gas forms the wind reprocessing the disc X-ray radiation. \cite{Fabrika2015} have found that the ULX optical counterparts have very hot winds resembling WNL-type stars and their optical spectra are also very similar to that of SS\,433. Using simple relations for the supercritical discs, they found that the optical luminosity of such discs is $L_V \propto \dot M_0^{9/4}$ because the stronger wind will reprocess more radiation emerging in the disc funnel.

We may conclude therefore that these two sources NGC\,5474 X-1 and M66 X-1 do belong to the ULX-type, and their high $L_{\mathrm{X}}/L_{\mathrm{opt}}$ ratios are determined as usual for ULXs X-ray luminosity, but a lower optical luminosity. This large ratio may be due to faint donor stars and it is quite possible that their optical emission might arise from these stars. Taking into account the colour and magnitude of the optical counterparts, they may represent F-G-type stars. Their optical luminosities might be also explained by lower (but nevertheless supercritical) mass accretion rates in these systems. They could be compared with the most dim ULXs in optical such as M81 X-6 or NGC\,1313 X-1 whose X-ray-to-optical luminosities are $L_{\mathrm{X}}/L_{\mathrm{opt}} \sim 4000$. 

\section*{Acknowledgements}
\noindent
The authors are grateful to D.\,Swartz and S. Balman for helpful comments regarding X-ray data of NGC\,5474 X-1. Our results are based on observations made with the NASA/ESA Hubble Space Telescope, obtained from the data archive at the Space Telescope Science Institute. STScI is operated by the Association of Universities for Research in Astronomy, Inc. under NASA contract NAS 5-26555. This research has made use of data obtained from the {\it Chandra} Data Archive and software provided by the {\it Chandra} X-ray Center (CXC) in the application package CIAO. This work made use of data supplied by the UK {\it Swift} Science Data Centre at the University of Leicester. This work is based on observations obtained with {\it XMM-Newton}, an ESA science mission with instruments and contributions directly funded by ESA Member States and NASA. The research was supported by the Russian RFBR grant 13-02-00885, the Program for Leading Scientific Schools of Russia N\,2043.2014.2, the Russian Scientific Foundation (grant N\,14-50-00043). SA acknowledge support from the Scientific and Technological Research Council of Turkey (TUBITAK) through project no. 113F039. SF and AV acknowledge support of the Russian Government Program of Competitive Growth of Kazan Federal University.

\label{lastpage}


\begin{thebibliography}{99}
\bibitem[\protect\citeauthoryear{Abolmasov et al.}{2007}]{Abolmasov2007a} Abolmasov P., Fabrika S., Sholukhova O., Afanasiev V., 2007a, Astrophys. Bull., 62, 36 
\bibitem[\protect\citeauthoryear{Abolmasov et al.}{2007}]{Abolmasov2007b} Abolmasov P.~K., Swartz D.~A., Fabrika S., Ghosh K.~K., Sholukhova O., Tennant A.~F., 2007b, ApJ, 668, 124 
\bibitem[\protect\citeauthoryear{Afanasiev \& Moiseev}{2005}]{Afanasiev2005} Afanasiev V.~L., Moiseev A.~V., 2005, Astron. Lett., 31, 194 
\bibitem[\protect\citeauthoryear{Aird et al.}{2010}]{Aird2010} Aird J., et al., 2010, MNRAS, 401, 2531 
\bibitem[\protect\citeauthoryear{Caballero-Garc{\'{\i}}a \& Fabian}{2010}]{Caballero2010} Caballero-Garc{\'{\i}}a M.~D., Fabian A.~C., 2010, MNRAS, 402, 2559 
\bibitem[\protect\citeauthoryear{Colbert \& Mushotzky}{1999}]{Colbert1999} Colbert E.~J.~M., Mushotzky R.~F., 1999, ApJ, 519, 89 
\bibitem[\protect\citeauthoryear{Della Ceca et al.}{2015}]{Della Ceca2015} Della Ceca R., et al., 2015, MNRAS, 447, 3227 
\bibitem[\protect\citeauthoryear{Drozdovsky \& Karachentsev}{2000}]{Drozdovsky2000} Drozdovsky I.~O., Karachentsev I.~D., 2000, A\&AS, 142, 425 
\bibitem[\protect\citeauthoryear{Fabrika}{1997}]{Fabrika1997} Fabrika S.~N., 1997, Ap\&SS, 252, 439 
\bibitem[\protect\citeauthoryear{Fabrika}{2004}]{Fabrika2004} Fabrika S., 2004, Astrophys. Space Phys. Rev., 12, 1 
\bibitem[\protect\citeauthoryear{Fabrika et al.}{2015}]{Fabrika2015} Fabrika S., Ueda Y., Vinokurov A., Sholukhova O., Shidatsu M., 2015, Nature Phys., 11, 551 
\bibitem[\protect\citeauthoryear{Feng \& Soria}{2011}]{Feng2011} Feng H., Soria R., 2011, New Astron. Rev., 55, 166 
\bibitem[\protect\citeauthoryear{Freitag, G{\"u}rkan,\& Rasio}{2006}]{Freitag2006} Freitag M., G{\"u}rkan M.~A., Rasio F.~A., 2006, MNRAS, 368, 141 
\bibitem[\protect\citeauthoryear{Gladstone, Roberts, \& Done}{2009}]{Gladstone2009} Gladstone J.~C., Roberts T.~P., Done C., 2009, MNRAS, 397, 1836 
\bibitem[\protect\citeauthoryear{Gladstone et al.}{2013}]{Gladstone2013} Gladstone J.~C., Copperwheat C., Heinke C.~O., Roberts T.~P., Cartwright T.~F., Levan A.~J., Goad M.~R., 2013, ApJS, 206, 14 
\bibitem[\protect\citeauthoryear{Gris{\'e} et al.}{2011}]{Grise2011} Gris{\'e} F., Kaaret P., Pakull M.~W., Motch C., 2011, ApJ, 734, 23 
\bibitem[\protect\citeauthoryear{Lee \& Jang}{2013}]{Lee2013} Lee M.~G., Jang I.~S., 2013, ApJ, 773, 13 
\bibitem[\protect\citeauthoryear{Liu, Bregman, \& Seitzer}{2004}]{Liu2004} Liu J.-F., Bregman J.~N., Seitzer P., 2004, ApJ, 602, 249 
\bibitem[\protect\citeauthoryear{Kaaret, Ward, \& Zezas}{2004}]{Kaaret2004} Kaaret P., Ward M.~J., Zezas A., 2004, MNRAS, 351, L83 
\bibitem[\protect\citeauthoryear{Kuntz et al.}{2005}]{Kuntz2005} Kuntz K.~D., Gruendl R.~A., Chu Y.-H., Chen C.-H.~R., Still M., Mukai K., Mushotzky R.~F., 2005, ApJ, 620, L31  
\bibitem[\protect\citeauthoryear{Madau \& Rees}{2001}]{Madau2001} Madau P., Rees M.~J., 2001, ApJ, 551, L27 
\bibitem[\protect\citeauthoryear{Motch et al.}{2014}]{Motch2014} Motch C., Pakull M.~W., Soria R., Gris{\'e} F., Pietrzy{\'n}ski G., 2014, Nature, 514, 198 
\bibitem[\protect\citeauthoryear{Medvedev \& Fabrika}{2010}]{Medvedev2010} Medvedev A., Fabrika S., 2010, MNRAS, 402, 479 
\bibitem[\protect\citeauthoryear{Ohsuga \& Mineshige}{2011}]{Ohsuga2011} Ohsuga K., Mineshige S., 2011, ApJ, 736, 2 
\bibitem[\protect\citeauthoryear{Pintore et al.}{2014}]{Pintore2014} Pintore F., Zampieri L., Wolter A., 
Belloni T., 2014, MNRAS, 439, 3461
\bibitem[\protect\citeauthoryear{Portegies Zwart et al.}{2004}]{Portegies2004} Portegies Zwart S.~F., Baumgardt H., Hut P., Makino J., McMillan S.~L.~W., 2004, Nature, 428, 724 
\bibitem[\protect\citeauthoryear{Poutanen et al.}{2007}]{Poutanen2007} Poutanen J., Lipunova G., Fabrika S., Butkevich A.~G., Abolmasov P., 2007, MNRAS, 377, 1187 
\bibitem[\protect\citeauthoryear{Poutanen et al.}{2013}]{Poutanen2013} Poutanen J., Fabrika S., Valeev A.~F., Sholukhova O., Greiner J., 2013, MNRAS, 432, 506 
\bibitem[\protect\citeauthoryear{Schaerer, Contini, \& Pindao}{1999}]{Schaerer1999} Schaerer D., Contini T., Pindao M., 1999, A\&AS, 136, 35
\bibitem[\protect\citeauthoryear{Schlegel, Finkbeiner, \& Davis}{1998}]{Schlegel1998} Schlegel D.~J., Finkbeiner D.~P., Davis M., 1998, ApJ, 500, 525 
\bibitem[\protect\citeauthoryear{Soria et al.}{2005}]{Soria2005} Soria R., Cropper M., Pakull M., Mushotzky R., Wu K., 2005, MNRAS, 356, 12 
\bibitem[\protect\citeauthoryear{Soria et al.}{2012}]{Soria2012} Soria R., Hakala P.~J., Hau G.~K.~T., Gladstone J.~C., Kong A.~K.~H., 2012, MNRAS, 420, 3599 
\bibitem[\protect\citeauthoryear{Stobbart, Roberts, \& Wilms}{2006}]{Stobbart2006} Stobbart A.-M., Roberts T.~P., Wilms J., 2006, MNRAS, 368, 397 
\bibitem[\protect\citeauthoryear{Sutton, Roberts, \& Middleton}{2013}]{Sutton2013} Sutton A.~D., Roberts T.~P., Middleton M.~J., 2013, MNRAS, 435, 1758 
\bibitem[\protect\citeauthoryear{Swartz, Tennant, \& Soria}{2009}]{Swartz2009} Swartz D.~A., Tennant A.~F., Soria R., 2009, ApJ, 703, 159 
\bibitem[\protect\citeauthoryear{Swartz et al.}{2011}]{Swartz2011} Swartz D.~A., Soria R., Tennant A.~F., Yukita M., 2011, ApJ, 741, 49 
\bibitem[\protect\citeauthoryear{Tao et al.}{2011}]{Tao2011} Tao L., Feng H., Gris{\'e} F., Kaaret P., 2011, ApJ, 737, 81 
\bibitem[\protect\citeauthoryear{Vinokurov, Fabrika, \& Atapin}{2013}]{Vinokurov2013} Vinokurov A., Fabrika S., Atapin K., 2013, Astrophys. Bull., 68, 139 
\bibitem[\protect\citeauthoryear{Yang, Feng, \& Kaaret}{2011}]{Yang2011} Yang L., Feng H., Kaaret P., 2011, ApJ, 733, 118 

\end{thebibliography}
\end{document}